# Path Loss and Directional Gain Measurements at 28 GHz for non-Line-of-Sight Coverage of Indoors with Corridors

Dmitry Chizhik, *Fellow, IEEE*, Jinfeng Du, *Member, IEEE,* Rodolfo Feick, *Senior Member, IEEE,* Mauricio Rodriguez, *Senior Member, IEEE,* Guillermo Castro, Reinaldo. A. Valenzuela, *Fellow, IEEE*

***Abstract*-Adequate coverage with high gain antennas is key to realizing the full promise of the bandwidth available at mm/cm wave bands. We report extensive indoor measurements at 28 GHz (1000 links, 9.9 million individual power measurements, 10 offices, 2 buildings), with/without line-of-sight (LOS) using a continuous wave channel sounder, with a $10^o$ spinning horn, capable of capturing a full azimuth scan every 200 ms, in up to 171 dB path loss to characterize coverage with 90% confidence level. The environment had prominent corridors and rooms, as opposed to open/mixed offices in latest 3GPP standards. Guiding in corridors leads to much lower RMS azimuth spread ($7^o$ median in corridor non-LOS vs. $42^o$ in 3GPP) and higher penetration loss into rooms and around corners (30-32 dB, some 12 dB more loss than 3GPP at 20 m non-LOS). Measured path gain in non-LOS is predicted by a mode-diffusion model with 3.9 dB RMS error. Scattering degraded azimuth gain by up to 4 dB in the corridor and 7 dB in rooms with 90% probability. Link simulations in a canonical building indicate every corridor needs an access point to provide 1 Gbps rate to adjoining rooms within 50 m using 400 MHz of bandwidth.**

## I. INTRODUCTION

The wide spectrum available at mm/cm wave bands promises very high rates. However, higher free-space, scattering, transmission and diffraction losses in these bands must be overcome. Increasing the carrier frequency by a factor of 14 (from 2 GHz to 28 GHz) increases path loss by about 23 dB, even in free space. Increasing bandwidth from 10 MHz to 1 GHz increases noise power by 20 dB. The overall SNR for a link with the same antenna gain and transmit power thus decreases by 43 dB. This reduction in SNR may be overcome through larger antenna gains. In specular channels, dominated by a single direction of departure/arrival, a properly oriented high directional gain antenna is effective in realizing its nominal gain, as measured in an anechoic chamber. Such an antenna may be steered electronically through a phased array, or, perhaps, through selecting best of fixed beams created through an RF beamformer, such as a Butler matrix. Once aimed correctly, such an antenna may use a single transmit/receive chain, which is economically attractive.

In practical deployments, however, scattering reduces the effective gain of a directional antenna and temporal fluctuation affects how often the beam needs to be re-aimed. A directional antenna receives certain directions with higher gain, while attenuating others. In the extreme case of equal power coming from all directions, received power gained through higher gain in some directions is offset by power lost through not receiving in other directions and all directional gain is lost. Full digital beamforming has the potential to realize full gain in a scattering environment at the cost of installing RF and digital chains behind each antenna element, which could be costly for numerically large arrays needed in mm/cm wave bands. It is of great interest to determine how much directional gain is retained by forming directional beams in real environments, which will be somewhere in between specular and fully scattered.

### A. *Previous work*

Numerous path gain measurement campaigns [1]-[18] have been conducted at mm/cm wave bands in various types of indoor environments (room-room, around the corner, corridor-room). In these measurements, one end of the link is kept at a fixed location while the other is moved to dozens of locations. Excess loss over free space for around-a-corner [8][15][16][27] or corridor-room [26] non-LOS (NLOS) links were reported to be 20 to 40 dB, depending on frequency and construction materials. Indoor LOS/NLOS measurements and models at 28 GHz and 73 GHz, using wide-band, directional sounders were reported in [5], for 5 transmit (Tx) locations and 33 receive (Rx) locations. There were 165 possible Tx-Rx location combinations, of which 48 had adequate signal-to-noise ratio (SNR) to report a measurement result. Measurements collected in LOS conditions, but with antennas not aimed at each other were labeled as “NLOS”. A “directional” path loss, i.e. specific to a particular combination of Tx-Rx antenna aim directions, as well as the “omnidirectional” path loss, comparable to that which may be measured at the same location by an omni antenna, were reported.

The latest standardized channel model, 3GPP 38.901 [20], extend to mm wave bands (up to 100 GHz) previously

Dmitry Chizhik, Jinfeng Du and Reinaldo A. Valenzuela are with the Bell Laboratories, Nokia, Holmdel, NJ.
Rodolfo Feick is with the Universidad Técnica Federico Santa María, (UTFSM), Valparaiso, Chile.
Mauricio Rodriguez and Guillermo Castro are from Escuela de Ingeniería Eléctrica de la Pontificia Universidad Católica de Valparaíso, Valparaíso, Chile.

developed models from 3GPP and other entities such as [10] and [19]. Its indoor channel model focues on open and mixed offices and shopping malls, which have very different propagation environment as compared to office buildings with corridors and rooms reported in this work. New findings will be compared against 3GPP recommendations in [20].

### B. Our contribution

Main contributions of this paper are:

- Characterization of an indoor environment with corridors in mm wave band, finding substantially higher path loss in NLOS than 3GPP indoor models designated for open/mixed offices.
- Direct measurement of antenna gain degradation by scattering from azimuthal scans with 1-degree granularity. Median azimuth spread in corridors, even for NLOS links, is found to be 7°, in contrast to 42° recommended by 3GPP in the absence of corridors.
- High confidence determination of both path gain and directional gain through massive data collection.

Measurements were collected in LOS conditions in corridors, as well as NLOS conditions in corridor-room links and around-corner links in corridor intersections.

Measured path gain dependence on distance in LOS and NLOS scenarios is represented in this work through the traditional [20] least mean squares fit of both distance slope and intercept, sometimes referred to as the "ABG model" (e.g. [20]), with RMS deviation of under 3.4 dB. In contrast, applying the standard 3GPP model (38.901) [20] results in 12 dB RMS error with respect to NLOS (corridor-room) measurements of path gain.

Current work demonstrates experimentally that, for buildings with corridors, corridors act as main signal conduits for reaching terminals in rooms. This is done through determining the dominant angles of arrivals via azimuth scan, and finding consistency of our path gain data with a previously developed theoretical model [22], based on guiding along the corridor, followed by coupling (scattering) towards the user terminal. The received signal azimuthal distribution in corridors is found to be narrow and aligned with corridors, even when the transmitter is in a room (NLOS), corresponding to 7° RMS azimuth spread. This is in contrast to 42° azimuth spread recommended by 3GPP [20] in NLOS. Again, we attribute this to guiding.

We find penetration loss into rooms to be 12 dB more severe at 28 GHz than previously measured at 2 GHz in the same corridor.

We also propose a new slope-intercept model for around corner propagation in intersecting corridors, which was not modeled previously for office buildings. This new model yields RMS fit errors less than 4 dB for multiple measurements conducted at two different buildings at 28 GHz and for the around corners data at 24 GHz [8].

Uncertainty in predicting average local path gain at a location is due to both location-location variability, known as shadow fading as well as uncertainty due to estimation of model parameters, i.e. slope and intercept, based on data fit. High reliability outage statistics (e.g., 90% coverage), require hundreds of links for a particular environment to make sure the model uncertainty is much less than the RMS spread in the data. In this work we aim to reliably (with 90% confidence) characterize coverage achievable with a beam steering base antenna, mounted in a typical office corridor. We answer these questions by collecting a statistically significant set of path gains and effective antenna gains.

Our current work derives its conclusions based on over 1000 continuous wave (CW) links measured at 28 GHz center frequency, each consisting of over 30 azimuth scans, collected from 10 offices in 2 buildings, containing a total of over 9.9 million individual power measurements (1341 links × 740 power samples/sec × 10 sec). Multiple measurement runs, each with about 100 links at ranges from 2 m to 70 m, allows for sampling of the environment for a variety of furniture arrangements and door/window locations.

The rest of the paper is organized as follows: measurement equipment and environment are described in Section II, path gain measurements and models for LOS, corridor-room and around the corner cases are described in Sections III, IV and V, respectively, and effective azimuth gain in Section VI. Rate estimates for corridor-mounted base stations covering rooms along corridors are presented in Section VII, followed by Conclusions in Section VIII.

## II. MEASUREMENT DESCRIPTION

### C. Measurement equipment

Characterizing propagation with high degree of statistical confidence at 28 GHz to distances of interest (about 100 m) and including NLOS condition require a very generous link budget and rapid data collection to get a substantial data set in reasonable time. To do so, we constructed a narrowband sounder, transmitting a 22 dBm CW tone at 28 GHz into a 55° (10 dBi) horn (all antenna beamwidths here refer to half-power beamwidths). Naturally the CW sounder gathers no information on delay properties. The signal received by the a 10° (24 dBi) horn, was amplified by several adjustable gain low-noise amplifiers (5 dB effective noise figure), mixed with a local oscillator, resulting in an IF signal centered at 100 MHz. The IF signal power, filtered with a 20 kHz-wide bandpass filter, was measured and converted to digital values with a power meter and stored on a computer. Both horn antennas were vertically polarized, with cross-pol isolation over 25 dB. The transmitter and the receiver have free running local oscillators, with a frequency accuracy better than $1{:}10^{-7}$.

To gain insight into azimuthal angular power distribution needed to characterize effective antenna gain we mounted the complete receiver, including the data acquisistion computer, on a rotating platform allowing a full angular scan every 200 ms, thus enabling commensurate tracking of temporal changes. Given the sampling rate of 740 power samples/sec, at 300 rpm, power measurement was captured every 2.5 degrees. This angular sampling is substantially finer than the 10° beamwidth of the spinning receive horn.

The system was calibrated in the lab and anechoic chamber to ensure measurement power accuracy. The full dynamic range of the receiver (from noise floor to 1 dB compression point) was found to be 50 dB, extensible to 75 dB using switchable receiver amplifiers. In combination with removable transmit attenuators (0-40 dB, used at very short ranges), measurable path loss allowing at least 10 dB SNR ranged from 61 dB (1 meter in free space) to 137 dB (e.g. 200 m range with 30 dB excess loss). Measurable path loss extends to 171 dB with directional antenna gains. This follows from 32 dBm EIRP, 24 dBi receive horn gain, 6 dB noise figure and 10 dB target SNR.

The rotating 10° horn receiver was tested in an open field at a range of 40 meters from the transmitter. The measured receive pattern is seen in Fig. 1 to be within 1 dB of that measured in the anechoic chamber, down to -40 dB.

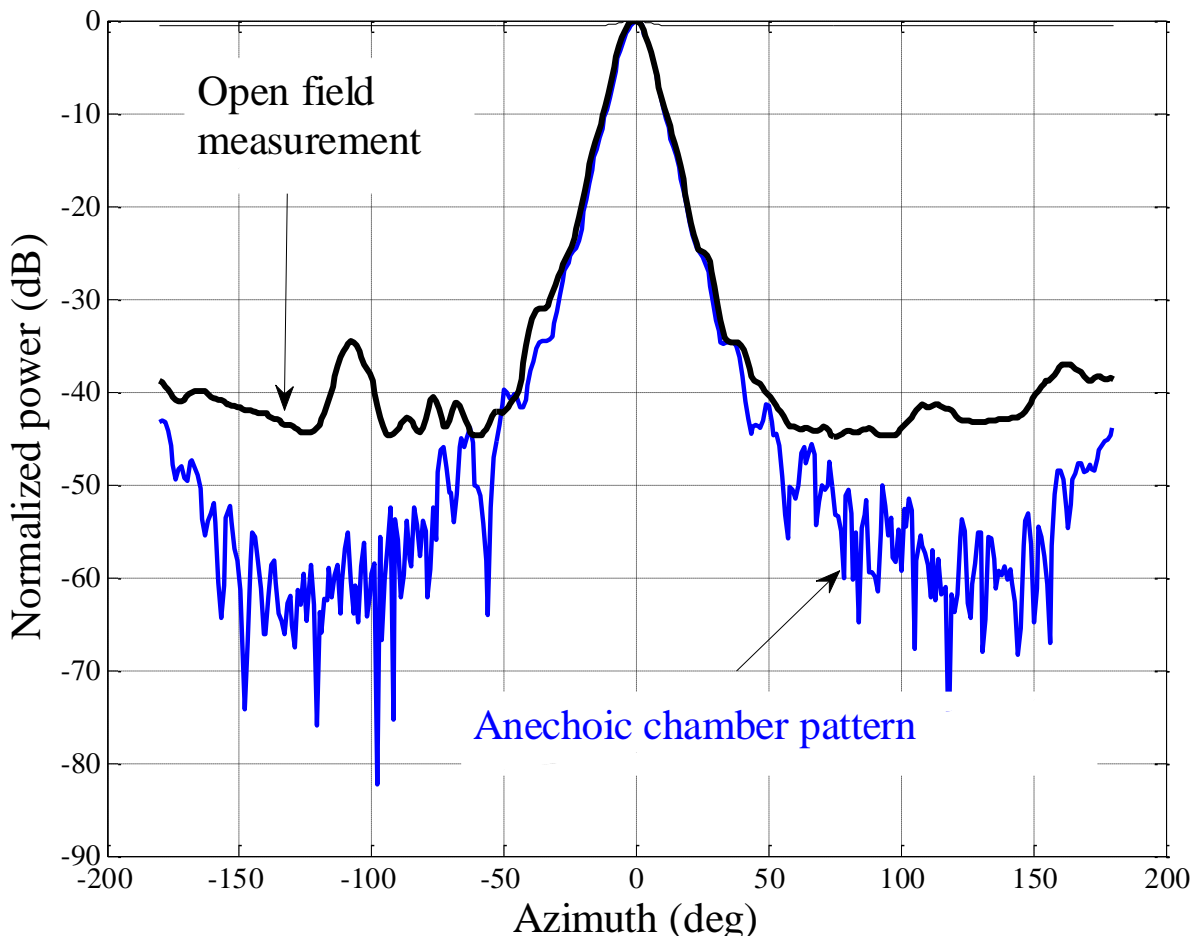

Figure 1. LOS measurements in open field match the anechoic measurements to -40 dB.

## D. *Measurement environment*

Measurements were performed in two buildings. The Bell Labs building in Crawford Hill, NJ has a 110 m long corridor, of 1.8 m width, lined with rooms. The interior walls are mostly plywood closets. LOS measurements were conducted in this corridor. In NLOS cases, one antenna was placed in a room while the other was placed in the corridor at different ranges, as illustrated in Fig. 2. The 55° transmit horn was manually aimed to get maximum received power. This is justified by finding that adding power from other transmit directions would not increase total power by more than 2 dB, i.e. the 55° beamwidth was sufficient to capture most power. NLOS measurements were also collected around the corner of an intersection of two such hallways. The building at Universidad Técnica Federico Santa María (UTFSM), Valparaíso, Chile also has long corridors, of 2.9 m width, lined with offices and laboratories. The interior walls are a mixture of sheetrock and concrete walls. The floor plan is similar in its essential aspects to the NJ building and we omit the layout for brevity.

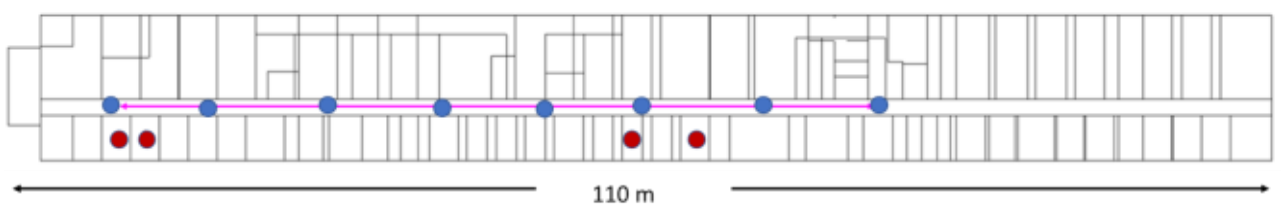

Figure 2. Corridor lined with rooms used for measurements in Crawford Hill, NJ. Schematic illustrates some of the corridor locations (blue dots) and room locations (red dots). LOS measurements were done between blue dots and NLOS (corridor-room) between red dots and blue dots.

## E. *Data processing for omni-directional path gain*

Average omnidirectional path gain $\langle P_{\text{omni}} \rangle$ for each measured link is computed by averaging power over all angular directions as follows.

The response of a receive antenna in direction $\phi_n$ is a circular convolution of the (unknown) complex directional channel response $h(\phi)$ and a complex antenna (field) response $a(\phi)$,

$$r(\varphi_n) = \int_0^{2\pi} d\varphi'\; h(\varphi')\; a(\varphi_n - \varphi') \qquad (1)$$

The received "power" is thus:

$$|r(\varphi_n)|^2 = \int_0^{2\pi}\int_0^{2\pi} d\varphi' d\varphi''\; h(\varphi') h^*(\varphi'') a(\varphi_n - \varphi')\; a^*(\varphi_n - \varphi''). \qquad (2)$$

The local average power over all azimuth directions follows as

$$\begin{aligned} P_{\text{all}} &= \frac{1}{2\pi}\int_0^{2\pi} d\varphi_n\; |r(\varphi_n)|^2 = \int_0^{2\pi} d\varphi' \int_0^{2\pi} d\varphi''\; h(\varphi')\; h^*(\varphi'') \\ &\times \frac{1}{2\pi}\int_0^{2\pi} d\varphi_n a(\varphi_n - \varphi')\; a^*(\varphi_n - \varphi''). \end{aligned} \qquad (3)$$

The expected value of the above average, under the uncorrelated scattering assumption:

$$\langle h(\phi') h^*(\phi'') \rangle = P(\phi')\delta(\phi' - \phi''), \qquad (4)$$

with the channel power angular spectrum $P(\phi) \equiv \langle |h(\phi)|^2 \rangle$ is thus:

$$\begin{aligned} \langle P_{\text{all}} \rangle &= \frac{1}{2\pi}\int_0^{2\pi} d\varphi'\; P(\varphi') \frac{1}{2\pi}\int_0^{2\pi} d\varphi_n |a(\varphi_n - \varphi')|^{\,2} \\ &= \frac{1}{2\pi}\int_0^{2\pi} d\varphi'\; P(\varphi') = \langle P_{\text{omni}} \rangle, \end{aligned} \qquad (5)$$

where the second equality follows from $\frac{1}{2\pi}\int_0^{2\pi} d\varphi_n |a(\varphi_n - \varphi')|^{\,2} = 1$ by definition of antenna directivity. The assumption (4) of uncorrelated arrivals from different directions [28] is satisfied even in the presence of a deterministic arrival (e.g. LOS), as arrivals from other directions are assumed independent and zero-mean.

Path gain $P_{\text{G}}$ is estimated from the azimuthal average (5) of received power by subtracting transmit power $P_{\text{T}}$, nominal transmit antenna gain $G_{\text{T}}$ and nominal elevation gain of the receive antenna $G_{\text{elev}}$:

$$P_{\mathrm{G}} = 10\log_{10}\left\langle P_{\mathrm{all}}\right\rangle - P_{\mathrm{T}} - G_{\mathrm{T}} - G_{\mathrm{elev}} \qquad (\mathrm{dB}) \qquad (6)$$

Both transmit antenna gain and receive elevation gain in (6) are assumed undegraded by scattering. The transmit antenna beamwidth (55°) employed in measurements is wider than the expected angle spread. Similar justification is made for the elevation gain, since the elevation angle spread has been reported as being small [20].

## III. PATH GAIN MEASUREMENTS IN CORRIDOR

Transmitter and receiver were placed in the same corridor, at ranges from 1 m to 76 m, every 0.9 m (3 feet) in NJ and from 2 m to 55 m in Chile (every 0.5 m), resulting in a total of 226 LOS links. Measured path gain dependence on distance $d$ in LOS, shown at the top of Fig. 3, was found to be well represented by the power law, with 90% confidence intervals [31], as indicated:

$$\begin{aligned} &P_{\mathrm{LOS}} = A + 10n\log_{10} d + N(0,\sigma), \\ &A = -62.6 \ +/- \ 0.7, \ n = -1.7 +/- \ 0.05, \ \sigma = 2.4 \text{ dB} \end{aligned} \qquad (7)$$

All distances are in meters and both $A$ and $n$ are fit to data. Since path gain (negative of path loss in dB) generally decreases with distance, its distance exponent $n$ is negative.

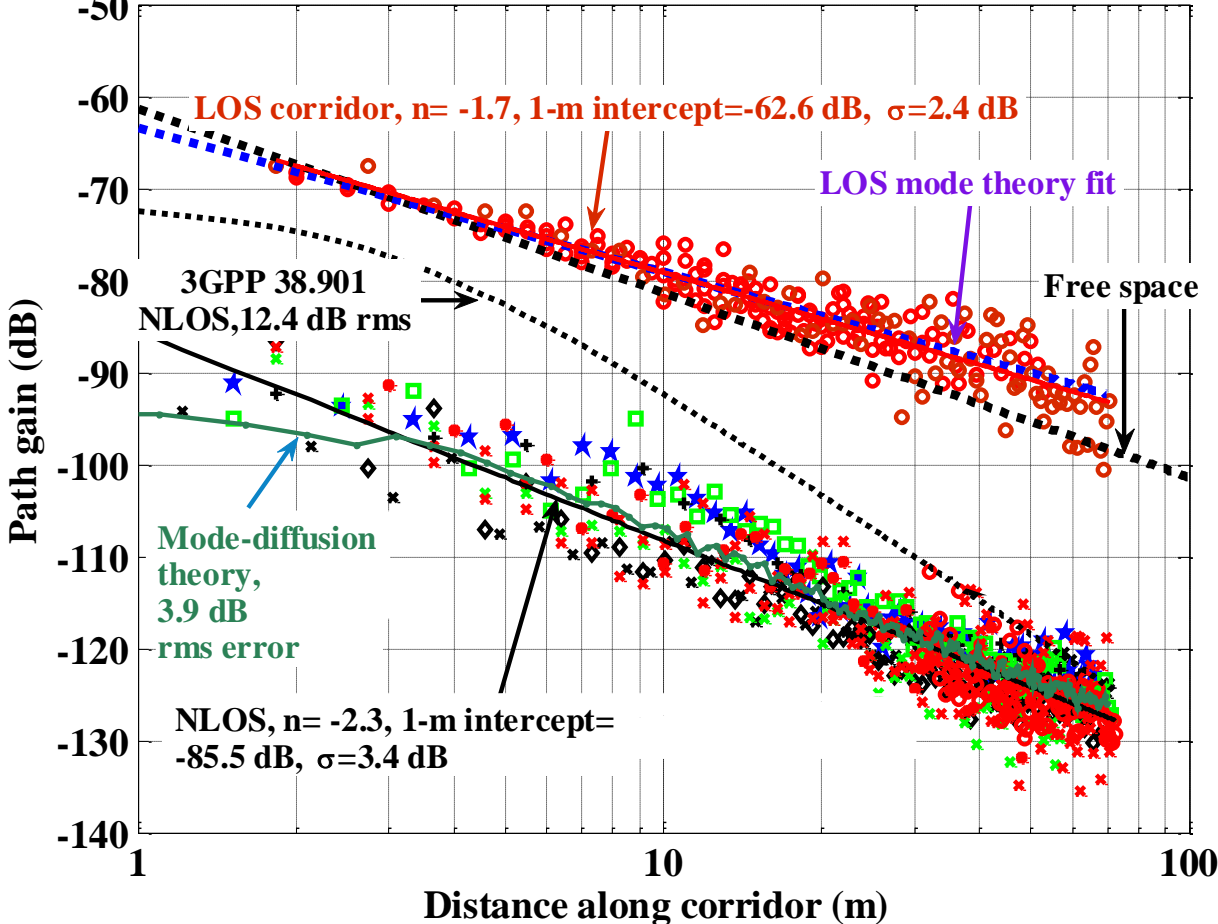

Figure 3. Path gain at 28 GHz measured in Crawford Hill, NJ and UTFSM, Chile. 226 LOS links and 697 corridor-room NLOS links (10 rooms). Different symbols indicate measurements in different rooms.

The deviation from the fit above was distributed within 0.5 dB of the lognormal distribution with 2.4 dB RMS. The 1-m intercept $A$ is within 1.2 dB from the corresponding Friis formula value at this frequency. The decay with distance is slower than the power of 2, expected in free space, consistent with guiding in the corridor (e.g. [10],[13]). This is similar to the LOS observations in the same corridor at 2 GHz [22], once the intercept is scaled with frequency $f$, as $f^2$. The walls are within 1 m of the antenna(s) placed in the corridor, with the attendant reflections sufficient to raise the received power (averaged over all azimuths) 2 to 3 dB above free space even at a range as short as 2.5 m, as may be verified by a simple sum of direct and reflected ray calculation.

## IV. CORRIDOR-ROOM PATH GAIN MEASUREMENTS

NLOS link measurements were made with the high directivity (spinning) horn in the corridor and the low directivity horn in different offices, and vice-versa, at ranges from 1.2 to 70 m for a total of 697 measured NLOS links, using 10 rooms (offices/labs). Each link measurement lasted for 10 seconds and consisted of over 30 full azimuthal scans, each with power measurements over 360 degrees, for a total of 7400 power measurements per link. Path gain data, collected in corresponding conditions in a University building at UTFSM in Chile, was found to be entirely consistent, in distance-dependence and RMS deviation, with the NJ data, despite apparent differences in interior wall composition and corridor width. Comparing the slope-intercept model derived from the NJ data to UTFSM data, as opposed to its own fit, led to an increase in RMS error of 0.2 dB. This was judged as small enough to allow the data to be combined. Measured NLOS path gain, shown in Fig. 3, was found to be well represented by the power law, with 90% confidence intervals [31], as indicated:

$$\begin{aligned} &P_{\mathrm{NLOS}} = A + 10n\log_{10} d + N(0,\sigma), \\ &A = -85.5 \ +/- \ 0.8, \ n = -2.3 +/- \ 0.06, \ \sigma = 3.4 \text{ dB} \end{aligned} \qquad (8)$$

Joint 90% model uncertainty [31] is under 1 dB for the measured range, and RMS model uncertainty is 0.24 dB, thanks to the large data set. The RMS deviation from the slope-intercept fit was 3.4 dB, distributed within 0.5 dB of the corresponding lognormal distribution, at outage levels from 5% to 95 %, shown in Fig. 4. Fixing the 1-m intercept at the Friis formula value of - 61.4 dB and adjusting only the distance exponent (that is, using the CI model [19]), increases the RMS error by 2.6 dB to 6 dB.

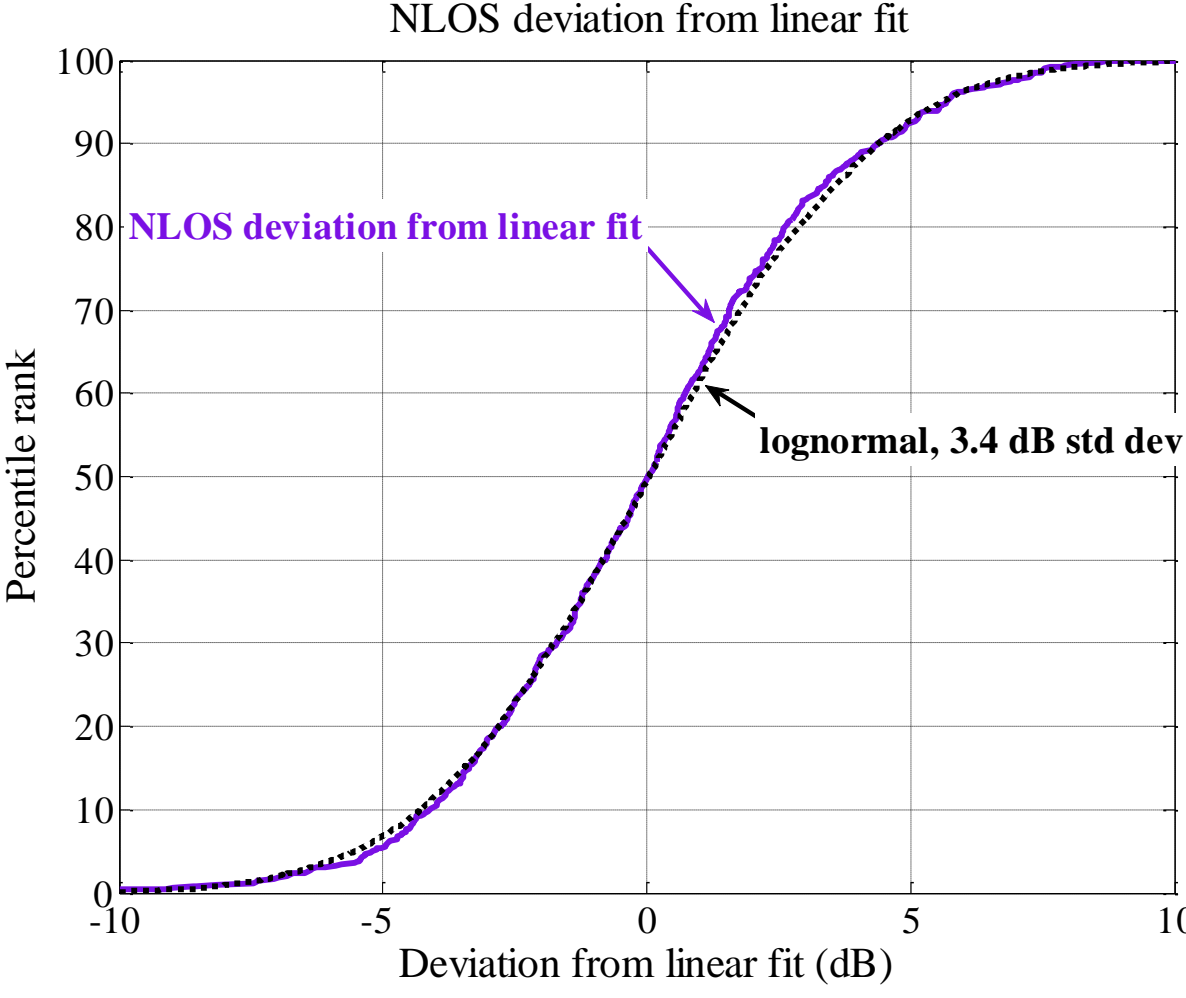

Figure 4. Distribution of deviation from linear fit in corridor-room NLOS ("shadow fading")

The excess loss of corridor-room NLOS relative to corridor LOS measurements at the same distance ranges from 20 to 40 dB, with a 32 dB median loss, in contrast to the corresponding 20 dB median loss found at 2 GHz [22]. This is in line with the approximately linear frequency dependence of the "office-to-corridor" excess loss reported in [26] over the frequency range of 2.45 GHz (20 dB loss) to 24 GHz (28 dB loss) based on a single office-to-corridor measurement route.

Notably the 3GPP [20] NLOS model for a more heterogeneous environment of office-office and open-cubicle links shows around 12 dB RMS deviation when compared to our findings. Measurement results in mmMagic [10] report lower losses but steeper exponents (3.5-3.9) than findings here. The lower path loss reported in [10] for corridor-room links may be due to lower-loss wall materials there (glass and plasterboard) as opposed to ones in the present study (plywood cabinets in NJ, or else, plasterboard interrupted every 10 m by concrete pillars in Chile).

A theoretical mode-diffusion model in [22] was found to provide predictions within 3.1 dB RMS error of path gain measurements at 2 GHz. In the model, the corridor is a lossy waveguide, and room interiors are diffusely scattering regions. Evaluating the same model at 28 GHz to compare against current 28 GHz data, resulted in RMS errors under 3.9 dB in both LOS and NLOS. Altering materials (e.g. smooth metal walls, vs. plywood walls) and corridor widths within the theoretical model exhibits a strong variation of overall losses, explaining the diverse findings here and in the references [10] [15] [17] [20] [26].

## V. PATH GAIN MODEL FOR INDOOR CORRIDOR-TO-CORRIDOR WITH TURN AROUND CORNERS

For buildings with intersecting corridors, NLOS transmission around corners suffers extra attenuation caused by scattering/diffraction as compared to LOS transmission along a single corridor. Measurements conducted at 24 GHz [8] [15] reported about 5 to 20 dB loss for turning around a corner. Measurements of co- and cross- polarization transmission in corridors at both 28 GHz and 73 GHz, each for 5 NLOS links, described in [18], reported RMS linear fitting errors ranging from 3.3 to 13.4 dB. Since traditional slope-intercept heuristic models based on Euclidean distance perform poorly in predicting path gain for corridor NLOS transmissions as reported in [15][18], it is desirable to develop new models to better predict the around-corner path gain for indoor buildings.

We illustrated in Fig. 5 a NLOS transmission link along the corridor passing around two corners, where the length of the three segments are $x_1$, $x_2$, and $x_3$ meters, respectively. For $d < x_1$, both terminals are in the same corridor in LOS conditions. For $x_1 < d < x_1 + x_2$, the signal will pass around one corner and for $d > x_1 + x_2$ the signal will pass around two corners. As the unwrapped distance $d$, i.e., the "Manhattan distance", between the two terminal increases, the propagation channel changes from LOS to NLOS one-corner, and to NLOS two-corner. Similar classification has also been adopted in [24][25] for channel modeling of propagation along urban street canyons.

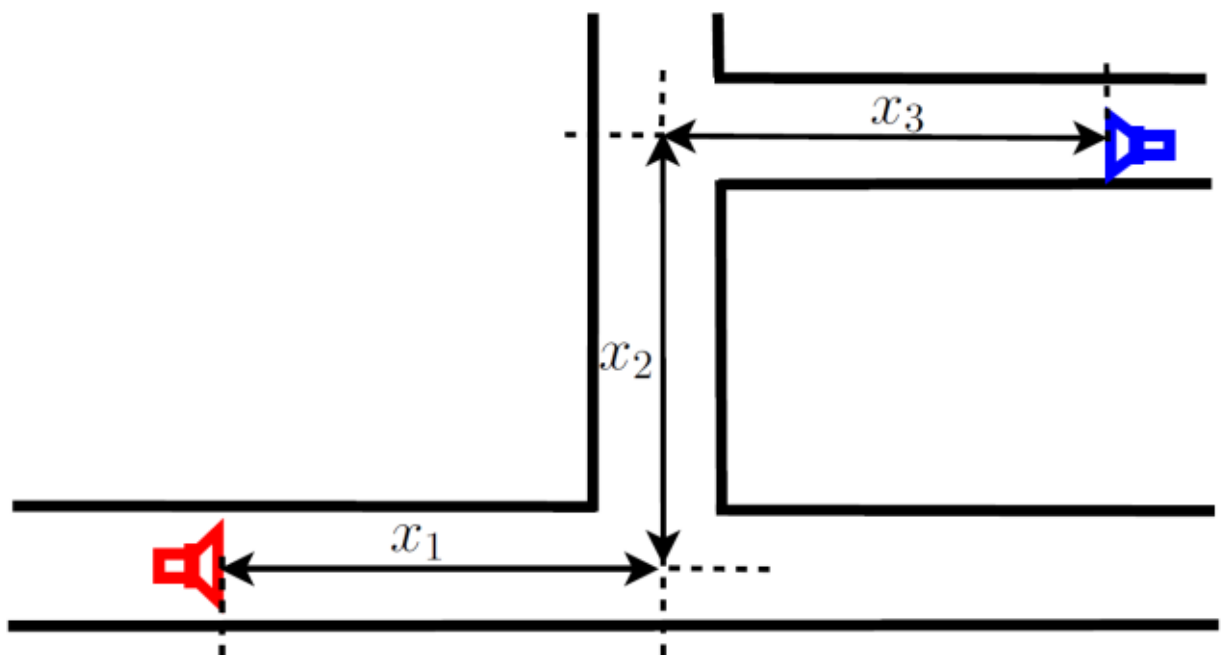

Figure 5. Illustration of an indoor transmission link along the corridor passing around two corners. One terminal (red) is placed at a fixed location inside the corridor, $x_1$ meters from a corner, and the other terminal (blue) moves away along the center line of corridors following a "Z-shaped" line trajectory.

Motivated by the clear waveguiding effect observed in the corridor LOS measurement, we propose a single-slope segment-wise path gain model for LOS/NLOS transmission along corridors, where the signal beyond each corner will be regarded as emitted from a new source located at the corner whose power is reduced by $S$ dB caused by diffraction/scattering around the corner. That is, the path gain for corridor transmission as a function of the traveled distance $d$ along the route is modelled as

$$P_{hallway}(d) = \begin{cases} P_1 + 10\,n\log(d), & 0 < d \le x_1; \\ P_1 - S + 10\,n\log(x_1(d - x_1)), & x_1 + w/2 \le d \le x_1 + x_2; \\ P_1 - 2S + 10\,n\log(x_1 x_2 (d - x_1 - x_2)), & d \ge x_1 + x_2 + w/2; \end{cases} \quad (9)$$

where $P_1$ is the intercept chosen to be the free space path gain at one meter, $n$ is the path-gain exponent, $S$ is the around-corner loss per turn, and $w$ is the width of the corridor. For $x_1 < d < x_1 + w/2$, the path gain is obtained by interpolation between $P_{hallway}(x_1)$ and $P_{hallway}(x_1 + w/2)$. Similarly, interpolation should be applied for $x_1 + x_2 < d < x_1 + x_2 + w/2$.

To validate the proposed model (9), measurements were done inside buildings with corridors and perpendicular intersections where the transmitter moves meter-by-meter along the center line of corridors and the receiver with a spinning horn is placed at a fixed location inside the corridor. Two data sets (HOH-1 and HOH-2) consisting of 119 links including over 880,000 individual power mesurements, were collected in Holmdel, NJ, with the receiver located at different ends of a "Z-shaped" line trajectory. Three data sets consisting of 299 links were collected in Valparaíso, Chile, along a "L-shaped" corridor with the receiver located at $x_1$=17, 26, and 37 meters, respectively, from the corner (labelled as USM-1, USM-2, and USM-3, respectively). By fitting the model (9) against all five data sets, we obtain $n$ = -1.81, around-corner turn loss $S$ of 18.7 dB, and RMS error of 3.0 dB. The measured data as well as the curves of the model stated in (9) using the above fitted parameters are presented in Fig. 6.

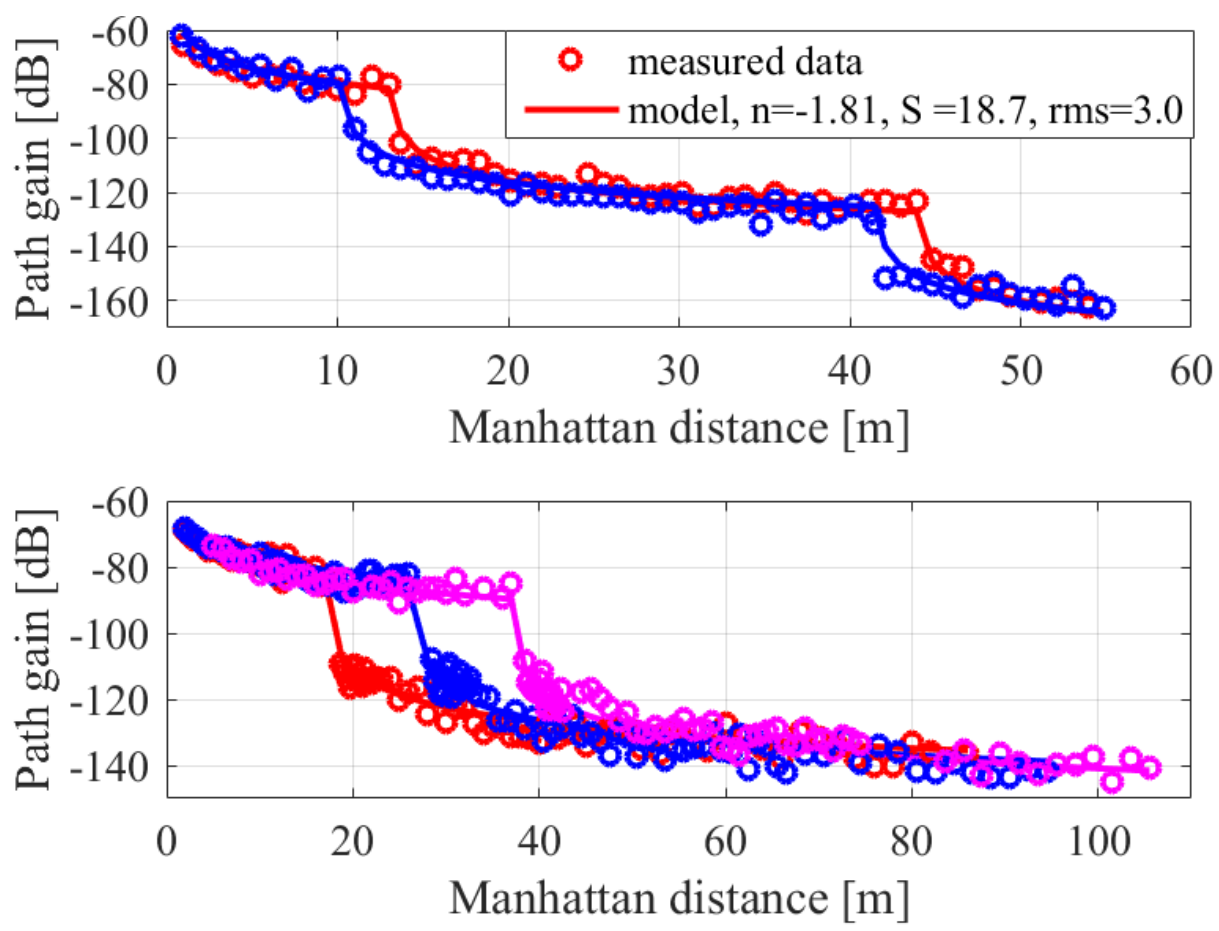

Figure 6. The proposed single-slope path gain model yields slope of $n$ = -1.81, turn loss of $S$ = 18.7dB, and RMS error of 3.0 dB when fitting to data from 418 measurement places: (upper) HOH two-turn data with 119 points; (lower) USM one-turn data with 299 points.

The proposed single-slope model is very robust. Firstly, despite the changes of turning distances, number of corners, and buildings, fitting each of the five data sets separately yield minor change of the slope and turn loss, as listed in Table I. Secondly, allowing different slopes and different corner losses for each segment can only reduce the RMS error by 0.2 dB, which justifies our choice of single-slope and common turn loss in model (9). Thirdly, the proposed model works equally well when tested against other measurement data sets. Swapping antenna locations, i.e., fixing the transmitter and moving the spinning-horn receiver to 99 measurement places around one corner (with $x_1$=31 m), resulted in a RMS error of 3.9 dB instead of 3.0 dB, when compared against the model (9) using the same fitting parameters ($n$=-1.81 and $S$=18.7) obtained from the combined five data sets. We also applied the model to the 24 GHz around-two-corner data set reported by Mehmood et al. [8]. By choosing the slope and intercept from the line-fitting to the LOS segment (i.e., before the first corner) and adjusting the fixed-turn-loss by $f$ (i.e, $S$ = 18 dB for 24 GHz as compared to $S$ = 18.7 dB obtained from our 28 GHz data), we obtain a less than 4 dB RMS error.

It is also interesting to compare our model (9) with other models regarding its accuracy. By introducing a fixed loss per turn, a simple model using free space formula for the route length (i.e., Manhattan distance) plus 30 dB loss per corner gives RMS error of 5.4 dB, as compared to RMS error of 12.8 dB obtained using a slope-intercept line fit against Euclidean distance. We note that the slope we obtained here, close to that of free space, is significantly smaller than the around-corner NLOS slope at 24 GHz reported in [15] where conventional slope-intercept model was fitted using Euclidean distance rather than Manhattan distance, but significantly larger than that obtained in [16], which was attributed to strong waveguiding effect. In [24] diffraction inspired path gain models are recommended for outdoor urban street canyon NLOS transmission at frequencies ranging from 430 MHz to 4860 MHz around one and two corners, respectively, where both terminals are below clutter.

Table I: fitting parameters of model (8) against measured data sets

| Data sets | Corners & lengths [m] | Corridor width | No. of links | Exponent $n$ | Corner loss $S$ [dB] | RMS error [dB] |
|---|---|---|---|---|---|---|
| HOH-1 | 2 corners, $x_1$=13, $x_2$=31 | 1.8 m | 58 | -1.79 | 17.8 | 2.4 |
| HOH-2 | 2 corners, $x_1$=10, $x_2$=31 | 1.8 m | 61 | -1.89 | 17.8 | 3.1 |
| USM-1 | 1 corner, $x_1$=17 | 2.9 m | 92 | -1.81 | 20.0 | 2.7 |
| USM-2 | 1 corner, $x_1$=26 | 2.9 m | 117 | -1.82 | 19.0 | 3.1 |
| USM-3 | 1 corner, $x_1$=37 | 2.9 m | 90 | -1.78 | 17.9 | 2.7 |
| **All data sets** | **--** | **--** | **418** | **-1.81** | **18.7** | **3.0** |

We extrapolate the street-canyon 1- and 2-turn NLOS models in [24] to indoor corridor transmissions and to 28 GHz, and reformulate a diffraction inspired indoor model as follows

$$P_{hallway}(d) = \begin{cases} P_1 + 10\,n\log(d), & 0 < d \le x_1; \\ P_1 - S_d + 5n\log(x_1(d - x_1)d), & x_1 + w/2 \le d \le x_1 + x_2; \\ P_1 - 2S_d + 5n\log(x_1 x_2 (d - x_1 - x_2)d), & d \ge x_1 + x_2 + w/2; \end{cases} \quad (10)$$

where $S_d$ is the diffraction loss for turning around the corner. Note that in the outdoor street canyon model [24] different loss parameters were obtained for the first and the second corners, and we assume the same loss by turning around a corner, without distinguishing if it is the first or the second corner. After fitting the diffraction inspired model (10) against the five data sets, we obtain $n$=-1.94, diffraction loss $S_d$=24 dB, and the corresponding RMS error is 3.6 dB, slightly worse than the 3.0 dB obtained using our model (9).

## VI. DIRECTIONAL GAIN MEASUREMENTS

High antenna gain is essential to compensate for the high propagation losses in mm/cm wave bands. As discussed in Sec. II, the effective pattern of an antenna is the convolution of its nominal pattern (as measured in an anechoic chamber) and the channel angular response (scattering pattern). This widens the effective antenna pattern, reducing its effective gain. In all cases, the azimuthal gain is defined as the ratio of the maximum power to average power over all azimuth angles:

$$\text{Azimuth gain} = \frac{\max_{\varphi} P(\varphi)}{\left(1/2\pi\right)\int_0^{2\pi} d\varphi\; P(\varphi)} \quad (11)$$

The degradation of azimuthal gain under angle spread conditions is illustrated in Fig. 7 for 3 cases: specular (no spread) where the gain scales linearly with the effective aperture, Gaussian channel angular spectrum of 42° RMS width, consistent with median azimuth departure spread specified by the 3GPP 38.901 InH NLOS model [20], which is seen to saturate as the array length increases, and finally equal power from all directions where no gain is realized at all. When the angular spectrum instantiation is also subject to direction-dependent fading, additional (small) gains are achievable by selecting the direction with the highest power instantiation. Below we report the effective azimuthal gain measured in our indoor environments.

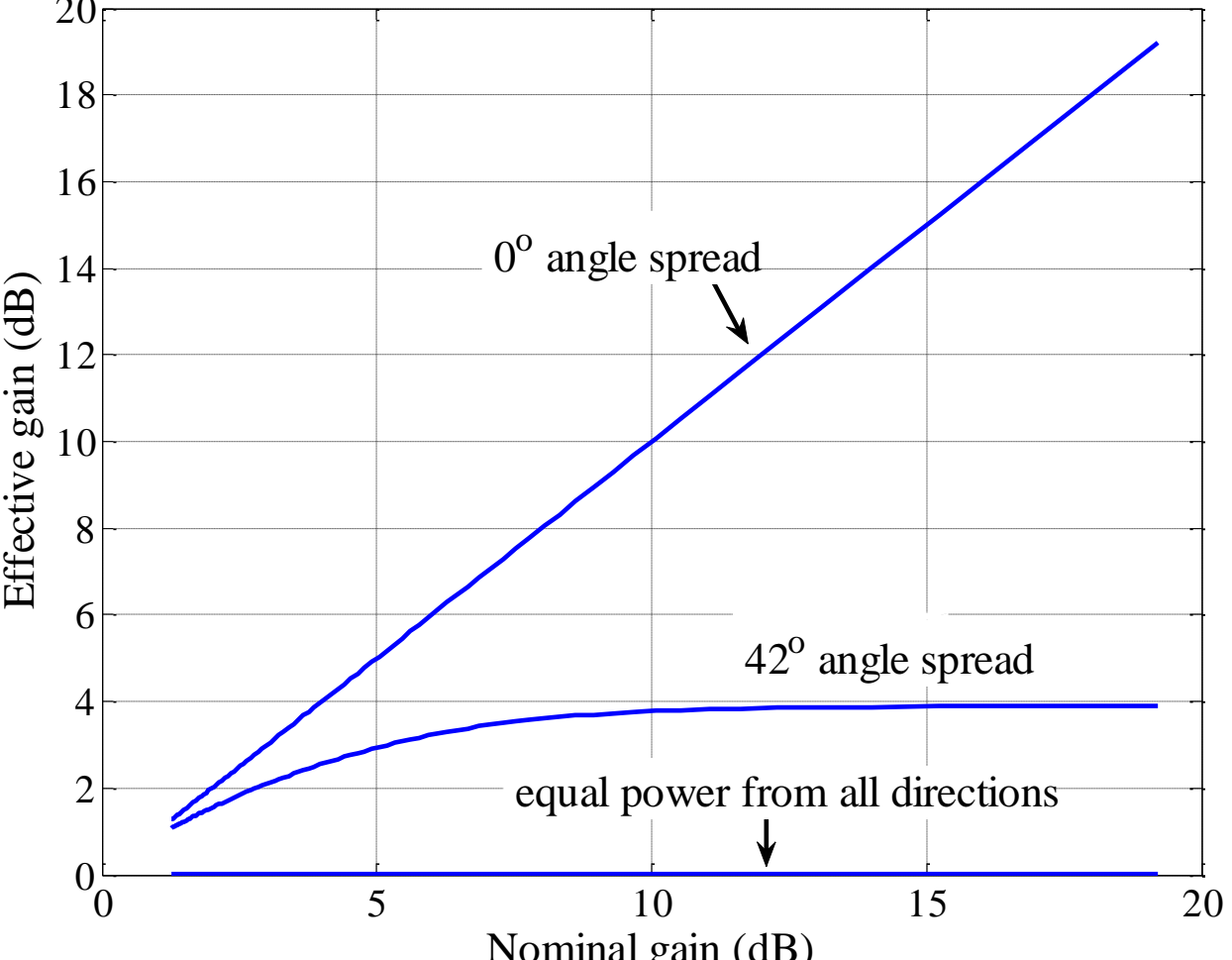

Figure 7. Effective directional antenna gain is limited by the channel angle spread.

Sample power angular spectra measured in NLOS, with spinning Rx in the corridor and in a room are illustrated in Fig. 8, where the greater scattering in the room is evident. In LOS (not shown), the measured patterns are narrow, as expected from guiding in a lossy waveguide.

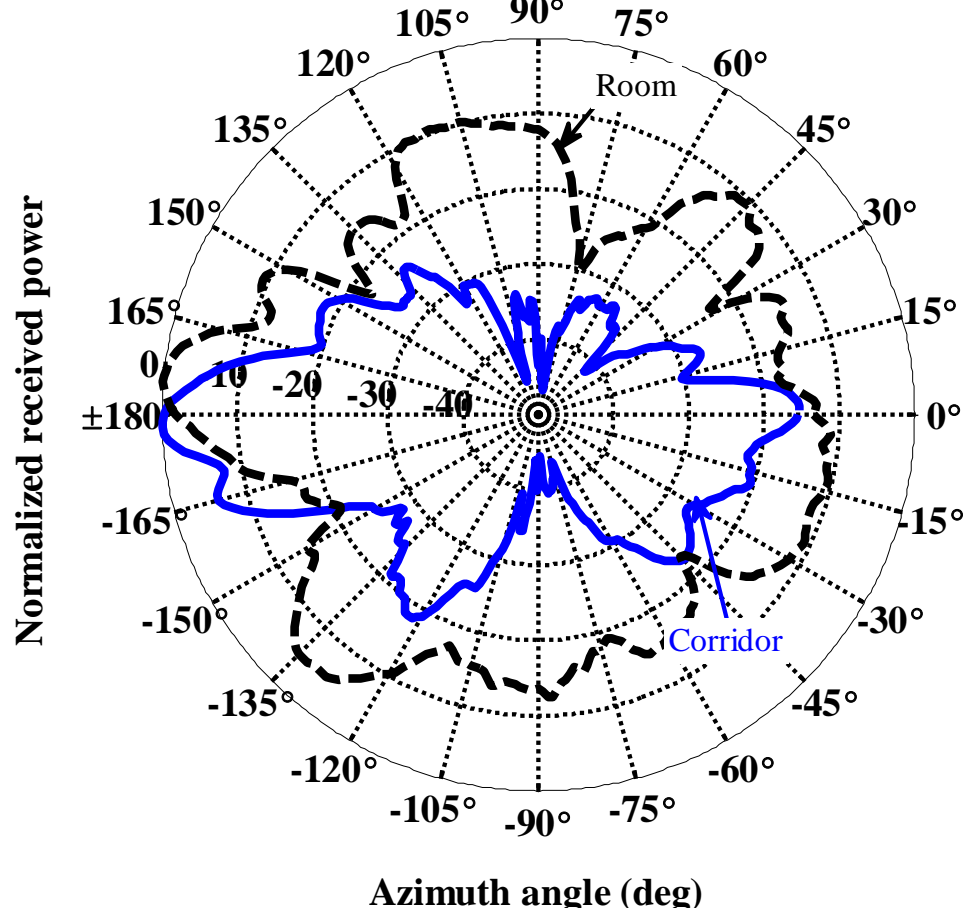

Figure 8. Sample measured power azimuth spectra in NLOS, with spinning Rx in corridor (blue) and room (black)

The empirical azimuthal gain distributions are shown in Fig. 9, with grey regions around a cumulative distribution function indicating 90% confidence bands [29]. The $10^{th}$ percentile azimuthal gain degradation from the ideal (14.5 dBi in anechoic chamber) is about 2.5 dB when both transmitter and receiver are in LOS corridor conditions. In NLOS conditions studied here, one end of the link is the room, the other in the corridor. The corresponding $10^{th}$ percentile azimuth gain degradation is 7.0 dB when measured in the room and 4.5 dB in the corridor. The implication is that, as expected, a high gain antenna is much more effective in a non-cluttered environment. Gain degradation measured in the room in NLOS is seen in Fig. 9 to be within 1 dB of the degradation following the 3GPP 38.901 recommendation for indoor NLOS (InH) [20]. In Fig. 9 the gain distribution from [20] was obtained numerically by convolving the complex antenna pattern with complex channel instantiations consistent with [20], (median RMS azimuth spread of 42° at the base and 50° at the terminal. The lower edge of the 90% confidence band is roughly 1 dB smaller than the median. Hence, no more than 5.5 dB gain loss at 90% locations with 90% confidence.

Distribution of RMS azimuth spread, computed from azimuthal gain measurements in the corridor is compared against 3GPP 38.901 indoor NLOS recommendations [20] in Fig. 10. Collected measurements have a 7° median RMS azimuth spread, in sharp contrast to 42° in NLOS 3GPP model [20]. The lower spread found here is likely due to the relatively narrow azimuth spread in the corridor, even in NLOS, and is indicative of how different the corridor environment is from the open/mixed office cases addressed in [20]. This is consistent with corridor acting as a conduit for reaching the terminal in a room, as modeled in [22].

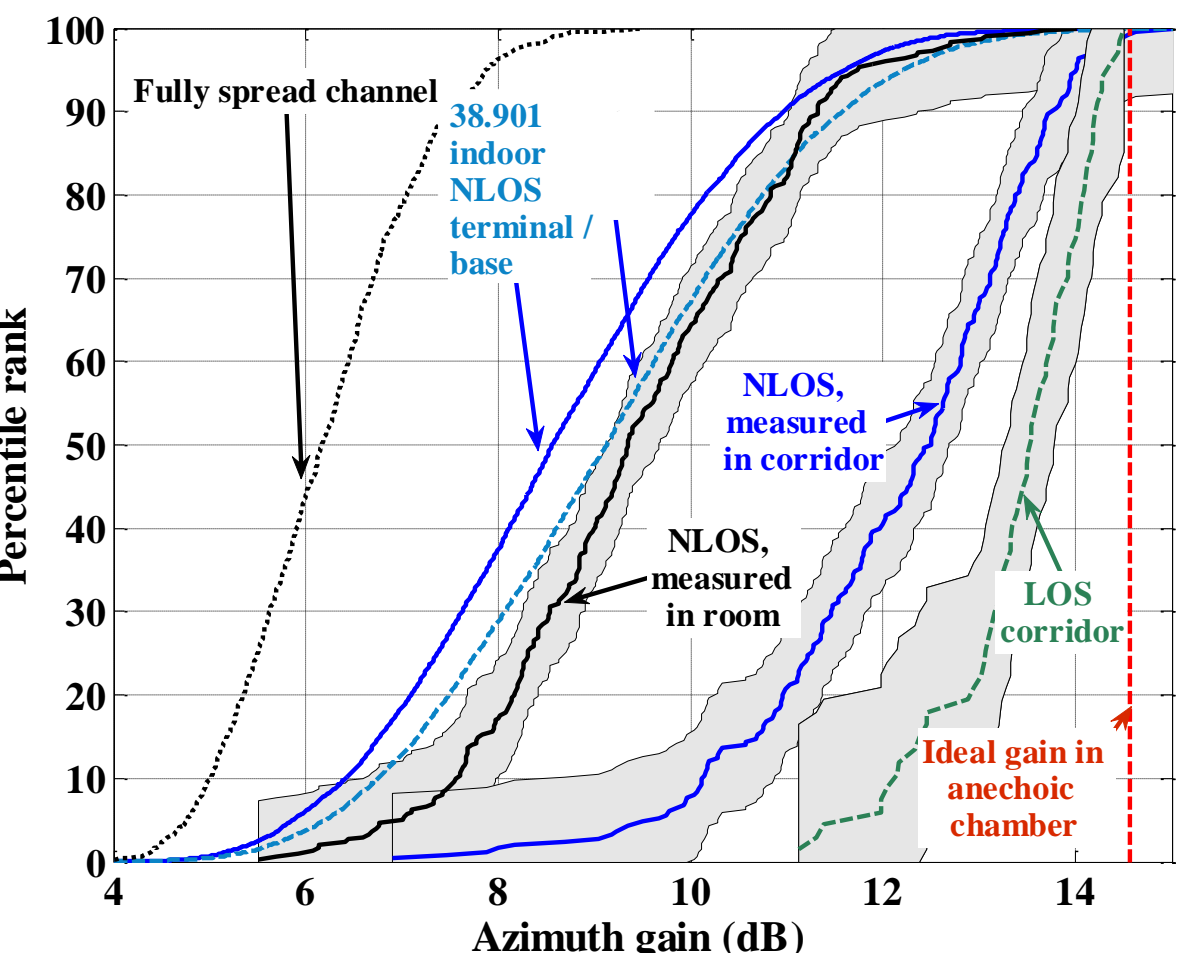

Figure 9. Effective azimuth gain distributions collected in a NJ office building. Gray regions indicate 90% confidence bands [29].

Observations of room power spectra, as in Fig. 8, showed that, for 75% of links, peak arrivals inside the room fell within a 70° "sector" centered around the direction towards the ("hot") wall adjoining the corridor. The actual realization of the peak varies within this range as the source moved along the corridor. Naturally the realization of the peak also depended on the room. The distribution of power measured from every direction within the 70° "sector" in the room, normalized by the average power over all directions (within the 70° "sector") is seen in Fig. 11 to be within 0.5 dB of the exponential distribution (Rayleigh fading).

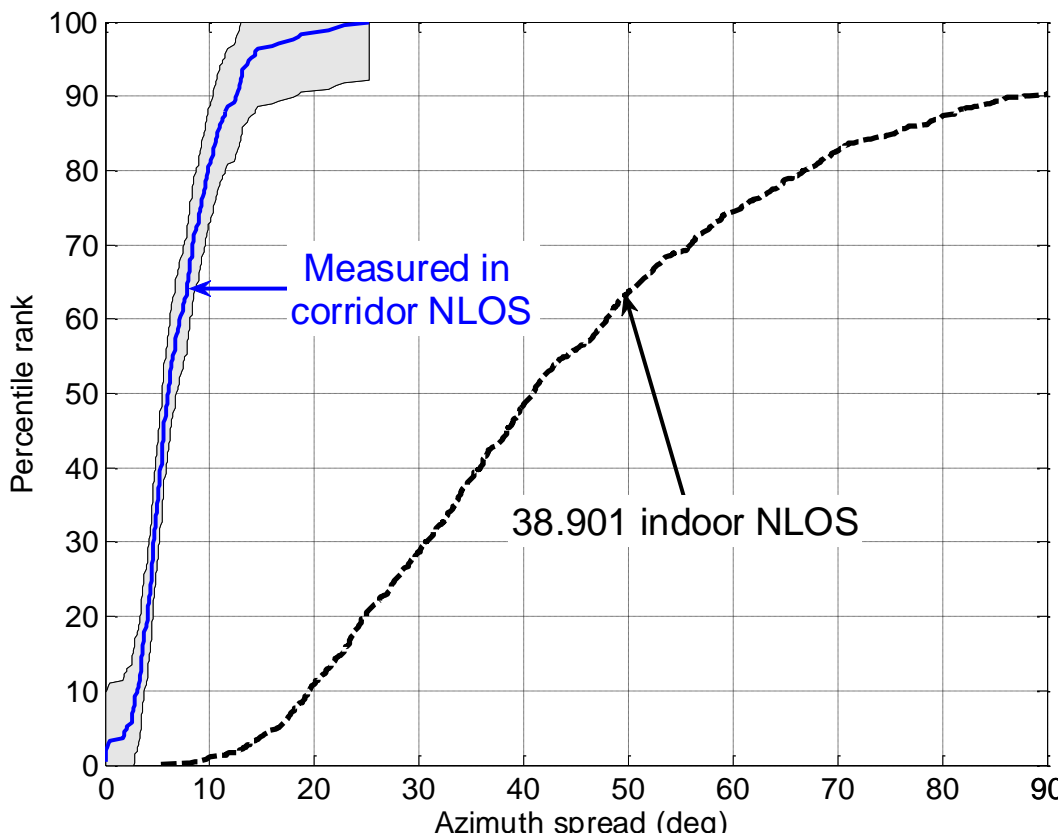

Figure 10. Effective azimuth spread computed from azimuthal scans in corridors for NLOS links in a NJ office building. Gray region indicates 90% confidence bands [29].

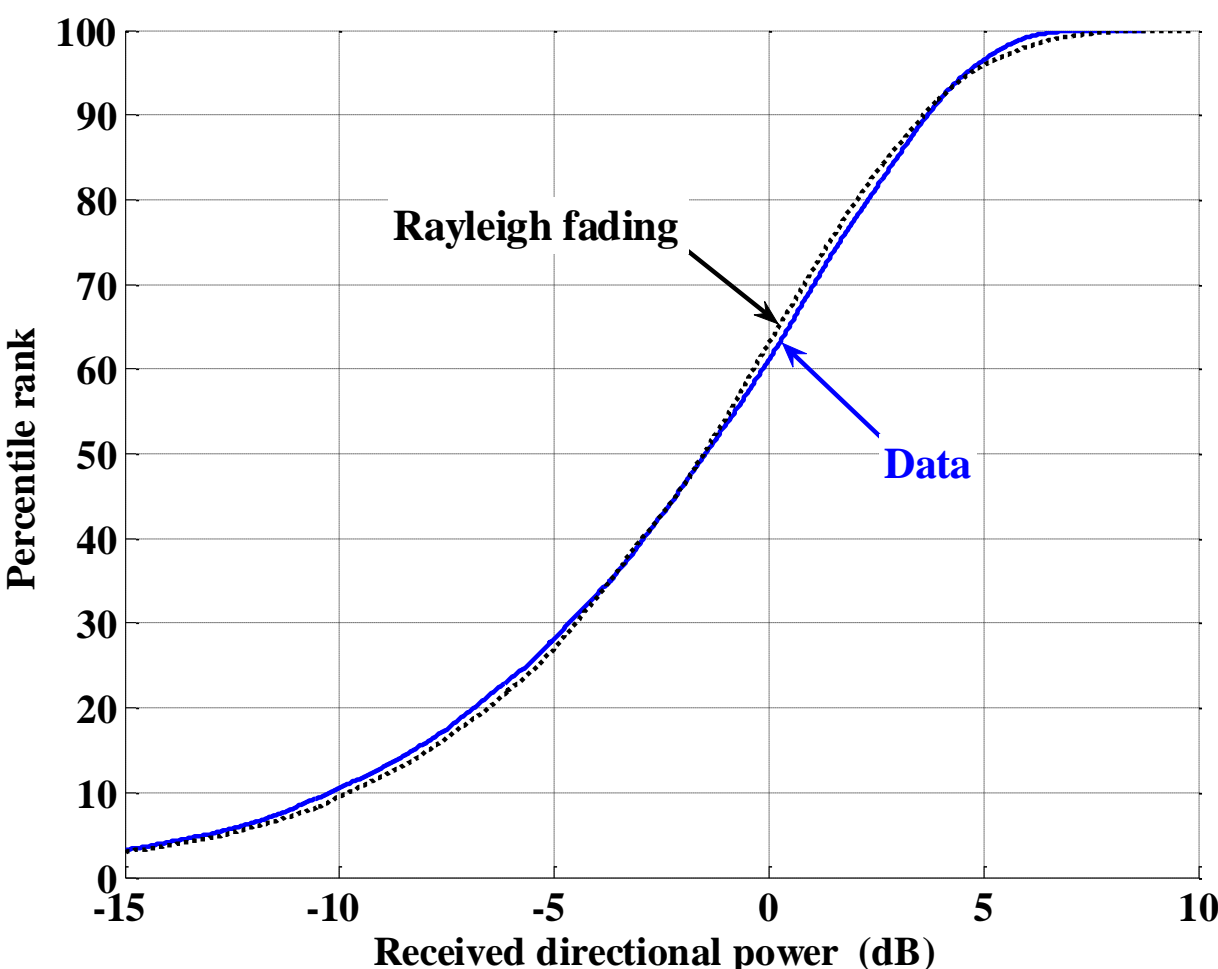

Figure 11. Distribution of deviation of received power in various directions relative to local mean power inside an office.

Choosing best aiming direction is beneficial even in a fully scattering channel, where power versus angle is constant *on average* yet the per instantiation power is Rayleigh distributed [28]. For reference, such a channel is simulated here by generating a dense grid in azimuth of incoming plane waves, of equal average power, but scaled by a complex Gaussian random realization, *CN*(0,1), representing diffuse (Rayleigh) scattering. The corresponding receive signal is obtained from (1) by convolving each such channel realization with the nominal complex pattern of our 24 dBi antenna, as measured in an anechoic chamber, and computing the effective azimuth gain as the ratio (11) of maximum to average over all aim angles. Repeating such a numerical experiment 10000 times results in a distribution of azimuth gain in a fully scattering channel, seen in the leftmost curve in Figure 9.

It may be observed that median effective azimuth gain measured in a room environment in NLOS conditions is only 3.5 dB above the gain available in a fully scattering channel. These angular diversity gains may be reduced in wide band channels, where best direction instantiation may vary across the band.

In the UTFSM building it was found that, while effective azimuth gains in LOS corridor are very similar to the results in NJ, in the case of NLOS, placing either the high gain spinning receiver in the room or in the corridor resulted in essentially indistinguishable distributions, seen in Fig. 12, with both well within joint 90% confidence band, shown in gray. In all cases, the low-gain transmit horn was manually aimed to ensure maximum received power. The 10% effective azimuth gain degradation was about 4.5 dB from the ideal gain, with rooms apparently less scattering. This may be due to thicker reinforced walls being used in earthquake-prone Chile, which constrains signal to enter through a wooden door or window. The corridor was (slightly) more scattering than in NJ, perhaps because it is wider, allowing for more angle spread due to reflections.

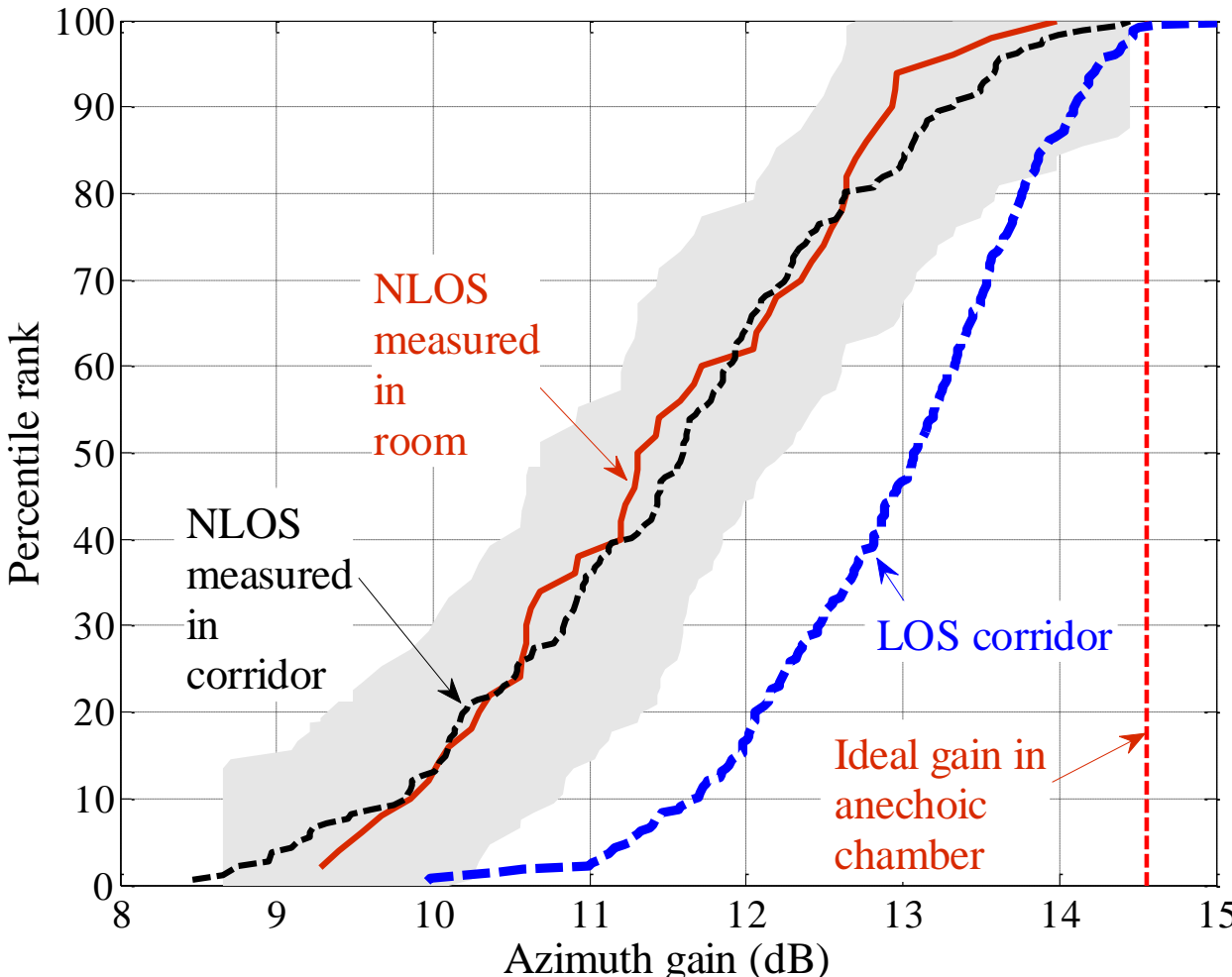

Figure 12. Effective azimuth gain distributions collected in UTFSM, Chile. 90% confidence band [29] plotted as grey region

## VII. Estimates of Achievable Indoor Rates

Downlink coverage and achievable rates are estimated as a function of range for the case of an access point (AP) in a corridor serving terminals inside rooms along that corridor for a canonical building, based on path gain models and gain degradation distributions obtained from the measurements. All links are thus NLOS, with propagation losses augmented at least by corridor-room penetration loss (~32 dB). The AP is assumed to have transmit power $P_T$=30 dBm, with (nominal) antenna gain of $G_{AP}$=24 dBi. The terminal antenna gain $G_{UE}$=5 dBi, with noise figure NF=9 dB and system bandwidth BW=400 MHz, and NLOS corridor-room path gain $P_G$ expression (8) were used. To cover 90% of locations at range $r$, a 90% margin $M$=6.7 dB was included, accounting for joint variation in shadow fading (3.4 dB RMS), directional gain loss (2 dB mean, 1.2 dB standard deviation), as shown in Figs. 4 and 9 (‘NLOS, measured in corridor’), and path gain model (8) uncertainty of 0.24 dB

$$SNR(r) = P_{\mathrm{T}} + G_{\mathrm{AP}} - P_{\mathrm{G}}(r) - M + G_{\mathrm{UE}} - Noise \quad (12)$$

The SNR is mapped directly to rate through the Shannon rate formula, $C = BW\log_2\left(1+SNR\right)$, (bps). Single spatial stream is assumed here. This is readily extendable to a dual pol, 2x2 MIMO link: given the high values of cross-polarization ratio reported in indoor channels (mean of 10 dB in [20]), a 2x2 cross-pol MIMO link is well approximated as 2 parallel channels with 10 dB isolation. The resulting rate is expected to be about double that of a single spatial stream. Even in the case of a single spatial stream per polarization, many antenna elements are needed to form the high directional gain represented by $G_{AP}$ in (12). An estimate of rates that can be achieved in a practical system would be implementation dependent and generally suffer from reduced effective SNR as well as effective bandwidth reduction due to overhead signaling, including pilots.

The resulting SNR and link Shannon rate are plotted vs. range in Figs. 13 and 14, respectively. These results may be contrasted against rates with an identically arranged 2 GHz, 10 MHz-bandwidth system with 5 dBi antennas at both ends and 30 dBm of transmit power. The 2 GHz path gain was obtained from the 28 GHz model found here after reducing the loss by 23 dB to account for $f^2$ and by 12 dB for the frequency-dependent difference in room penetration loss discussed above. The result is consistent with 2 GHz measurements [22]. The 5 dBi antenna gains at 2 GHz were taken as undegraded due to the already wide beams. It is seen that 1 Gbps link rates are achievable by this 28 GHz system at ranges of about 50 m and are about 10x the corresponding rates of the lower frequency system. The 40x increase in spectrum at 28 GHz versus 2 GHz, more than compensates for the reduced SNR and resulting lower spectral efficiencies at the higher frequency. Links with very high SNR would suffer additional impairments, such as phase and quantization noise, not considered here.

Performance estimates may be extended to other building layouts, including intersecting corridors, using formulas presented in previous sections, as appropriate. Likewise, mutual interference between multiple APs may be accounted for as well. While such results are particular to a specific building layout and AP deployment, it may be observed that, as discussed in Section V, terminals not adjoining the corridor where the AP is placed suffer a loss exceeding 25 dB for each corner turned, in addition to a median room penetration loss of 32 dB, reported in Section III. The combined excess loss of 57 dB ensures that signals that suffer both effects are below noise under above system power assumptions for ranges >10 m. Thus, base stations should be placed to illuminate every corridor to assure coverage of adjoining rooms.

As described, these results are based on sufficient empirical data to warrant 90% confidence in the model.

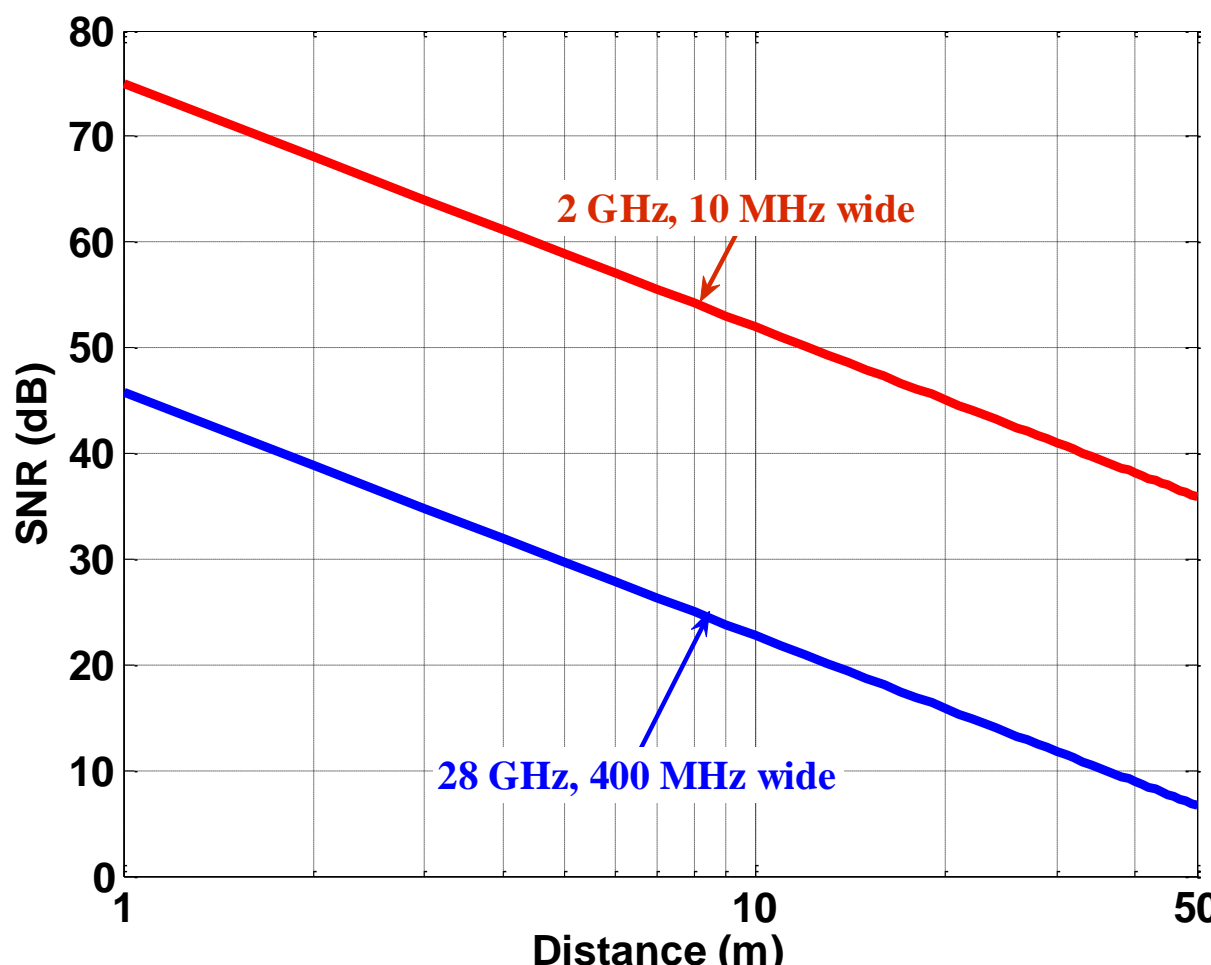

Figure 13. SNR dependence on distance in NLOS (corridor to room) 28 GHz, 400 MHz-wide, 24 dBi AP and 2 GHz, 10 MHz-wide, 5 dBi AP, both with 30 dBm transmit power. Includes 90% margins

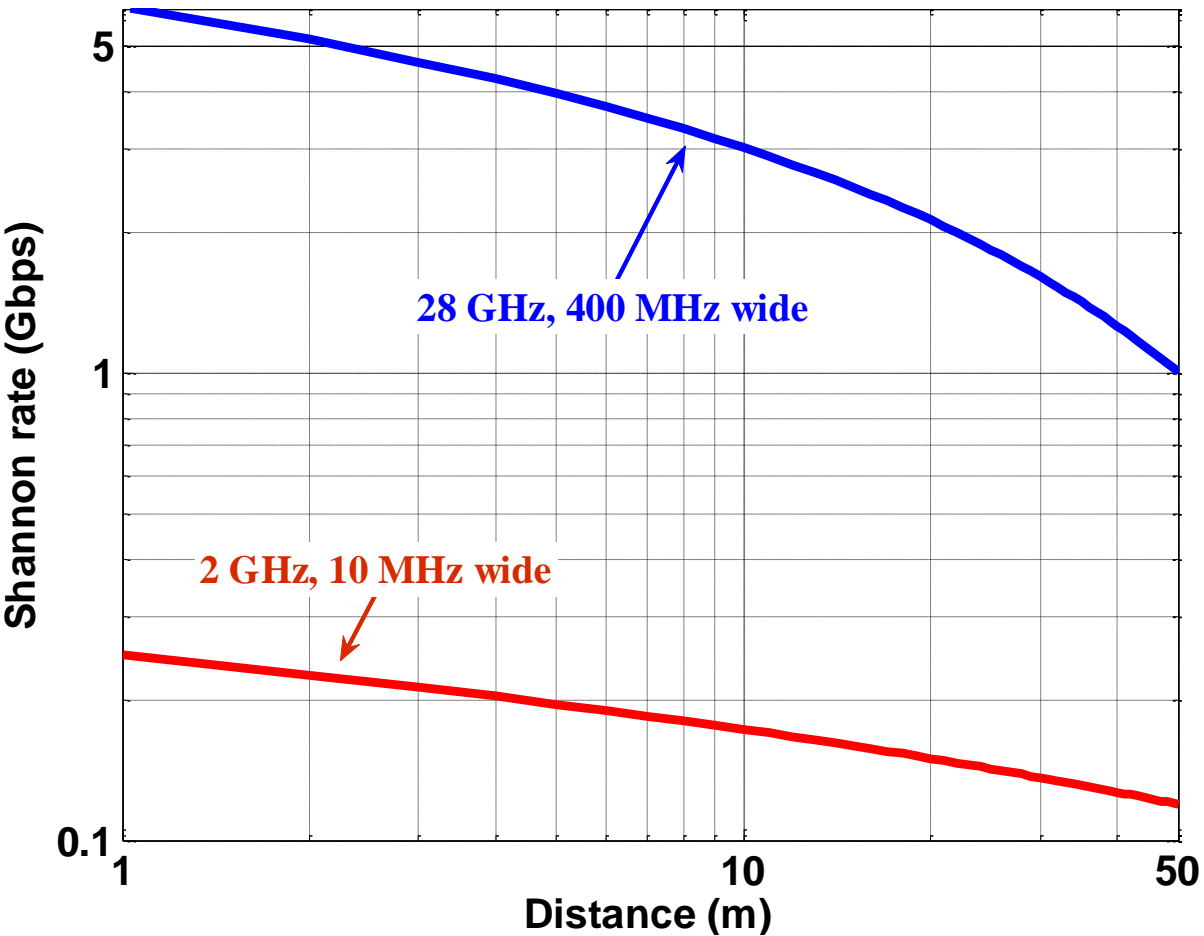

Figure 14. Shannon rate vs. distance for NLOS corridor-room links for 28 GHz, 400MHz-wide, 24 dBi AP and 2 GHz, 10 MHz-wide, 5 dBi AP. Transmit power is 30 dBm in both cases

## VIII. CONCLUSIONS

Extensive measurements (over 1000 links, each containing at least thirty 360-degree scans in azimuth, with over 9.9 million individual power measurements) of path gain and achievable azimuth gain were collected at 28 GHz in two office buildings using a specially constructed narrowband sounder that included a 10° receive horn rotating at speed up to 300 rpm. Path gain distance dependence was found to be well represented by slope-intercept models, with a distance exponent of -1.7 in LOS corridors and -2.3 in NLOS conditions. Average excess loss in NLOS, where impairment included both penetration into rooms as well as turning corners, was on the order of 30 dB per impairment, as compared to the 20 dB room penetration loss found at 2 GHz. Achievable azimuth gain, measured with a 10° horn placed in the corridor, was found to be degraded by no more than 5.5 dB in 90% of locations, with 90% confidence. Median measured RMS azimuth spread is $7^o$, in sharp contrast to $42^o$ in NLOS 3GPP model [20].

The findings are consistent with the view that in this environment, the waves are guided by the corridors and then scattered into the rooms via the nearest wall. Observed path gains are within 4 dB RMS error of the corresponding theory. The angular spectra in the corridor are found to be confined to a narrow angular range along the corridor, while the peak arrivals in the room are limited to a $70^o$ sector centered around the corridor wall containing the door, as opposed to the direction of the shortest path to the access point.

System performance simulations using the observed path gain and azimuth gain degradations in a representative indoor office building indicated that 1 Gbps sum rates are possible to deliver to 90% of locations at ranges reaching 50 meters in NLOS, provided that an access point was deployed in every corridor to allow adequate coverage in offices.

## Acknowledgment

The authors wish to acknowledge the support received from the Chilean Research Agency Conicyt, through research grants Conicyt-Basal FB0821, Project Conicyt-Redes 180144, Fondecyt Project 11171159, Project VRIEA-PUCV 039.430/2020 and 039.437/2020. Many thanks to Hector Carrasco, Leonardo Guerrero and Rene Pozo for designing and building the platform, Cuong Tran for essential diagnostics, and Huaiyuan Tu, Roberto Roberts and Alicia Musa for data collection software improvement and measurements.

## References

[1] C.R. Anderson, T.S. Rappaport, K. Bae, A. Verstak, N. Ramakrishnan. W.H. Tranter, C.A. Shaffer, L.T. Watson., "In-building wideband multipath characteristics at 2.5 & 60 GHz," in *Proc. IEEE Vehicular Technology Conference - VTC Fall*, 2002.

[2] H. Zhao, R. Mayzus, S. Sun, M. Samimi, J. K. Schulz; Y. Azar, K. Wang, G. N. Wong, F. Gutierrez, T. S. Rappaport, "28 GHz millimeter wave cellular communication measurements for reflection and penetration loss in and around buildings in New York city," in *Proc. IEEE International Conference on Communications (ICC)*, June 2013.

[3] S. Nie, G. R. MacCartney, Jr., S. S., and T. S. Rappaport, "72 GHz millimeter wave indoor measurements for wireless and backhaul communications," in *proc. IEEE 24th International Symposium on Personal Indoor and Mobile Radio Communications (PIMRC)*, Sept. 2013.

[4] A. Karttunen, K. Haneda, J. Järveläinen, J. Putkonen, "Polarisation characteristics of propagation paths in indoor 70 GHz channels," in *proc. 9th European Conference on Antennas and Propagation (EuCAP)*, Apr. 2015.

[5] G. R. Maccartney, T. S. Rappaport, S. Sun, S. Deng., "Indoor office wideband millimeter-wave propagation measurements and channel models at 28 and 73 GHz for ultra-dense 5G wireless networks," *IEEE Access*, vol. 3, pp. 2388–2424, Oct. 2015.

[6] S. Sun, G. R. MacCartney and T. S. Rappaport, "Millimeter-wave distance-dependent large-scale propagation measurements and path loss models for outdoor and indoor 5G systems," in *Proc. 10th European Conference on Antennas and Propagation (EuCAP)*, Davos, 2016

[7] O. H. Koymen, A. Partyka, S. Subramanian, J. Li., "Indoor mm-wave channel measurements: comparative study of 2.9 GHz and 29 GHz," in *proc. IEEE Global Communications Conference (GLOBECOM)*, San Diego, CA, Dec. 2015.

[8] R. Mehmood, J. W. Wallace and M. A. Jensen, "LOS and NLOS millimeter-wave MIMO measurements at 24 GHz in a hallway environment," in *proc. IEEE International Symposium on Antennas and Propagation (APSURSI)*, Fajardo, Jun. 2016.

[9] R. Mehmood, J. W. Wallace, W. Ahmad, Y. Yang, and M. A. Jensen, "A Comparison of 24 GHz and 2.55 GHz MIMO Measurements in Two Indoor Scenarios", in *proc. IEEE Antennas and Prop. Conf.*, July 2017.

[10] "Measurement Campaigns and Initial Channel Models for Preferred Suitable Frequency Ranges", mmMagic Project Report: "Millimetre-Wave Based Mobile Radio Access Network for Fifth Generation Integrated Communications (mmMAGIC), Deliverable D2.1, 2016.

[11] S. Deng, G. R. MacCartney and T. S. Rappaport, "Indoor and outdoor 5G diffraction measurements and models at 10, 20, and 26 GHz," in *proc. IEEE Global Communications Conference (GLOBECOM)*, Washington, DC, Dec. 2016.

[12] J. Ko, Y.-J. Cho, S. Hur, T. Kim, J. Park, A. F. Molisch, K. Haneda, M. Peter, D. Park, D.-H. Cho, "Millimeter-Wave Channel Measurements and Analysis for Statistical Spatial Channel Model in In-Building and Urban Environments at 28 GHz", *IEEE Trans. Wireless Communications*, vol. 16, Sept. 2017.

[13] V. Raghavan, A. Partyka, L. Akhoondzadeh-Asl, M. A. Tassoudji, O. H. Koymen and J. Sanelli "Millimeter wave channel measurement and implications for PHY layer design", *IEEE Trans. Antennas and Prop*, vol. 65, pp. 6521-6533, Dec. 2017.

[14] D. Chizhik, J. Du, G. Castro, M. Rodriguez, R. Feick, R.A. Valenzuela," Path Loss Measurements and Models at 28 GHz for 90% Indoor Coverage", *Proc. 12th European Conference on Antennas and Propagation (EuCAP)*, 2018.

[15] J. W. Wallace, W. Ahmad, Y. Yang, R. Mehmood and M. A. Jensen, "A Comparison of Indoor MIMO Measurements and Ray-Tracing at 24 and 2.55 GHz," *IEEE Transactions on Antennas and Propagation*, vol. 65, no. 12, pp. 6656-6668, Dec. 2017.

[16] S. Geng and P. Vainikainen, "Millimeter-Wave Propagation in Indoor Corridors," *IEEE Antennas and Wireless Propagation Letters*, vol. 8, pp. 1242-1245, 2009.

[17] F. Huang, L. Tian, Y. Zheng and J. Zhang, "Propagation Characteristics of Indoor Radio Channel from 3.5 GHz to 28 GHz," *in proc. IEEE 84th Vehicular Technology Conference (VTC-Fall)*, Sept. 2016.

[18] G. R. MacCartney, Jr., S. Deng, and T. S. Rappaport, "Indoor Office Plan Environment and Layout-Based MmWave Path Loss Models for 28 GHz and 73 GHz," in *proc. IEEE 83rd Vehicular Technology Conference Spring (VTC 2016-Spring)*, May 2016.

[19] "5G channel model for bands up to 100 GHz," 5GCM White paper, Dec. 2015. http://www.5gworkshops.com/5GCM.html

[20] "Study on channel model for frequencies from 0.5 to 100 GHz," 3GPP Technical Report TR 38.901 v14.1.1, Jul. 2017. http://www.3gpp.org/ftp/specs/archive/38_series/38.901/38901-e11.zip

[21] K. Haneda, L. Tian, H. Asplund, J. Li, Y. Wang, D. Steer, C. Li, T. Balercia, S. Lee, Y. Kim, A. Ghosh, T. Thomas, T. Nakamurai, Y. Kakishima, T. Imai, H. Papadopoulas, T. S. Rappaport, G. R. MacCartney, M. K. Samimi, S. Sun, O. Koymen, S. Hur, J. Park, J. Zhang, E. Mellios, A. F. Molisch, S. S. Ghassamzadeh, A. Ghosh "Indoor 5G 3GPP-like channel models for office and shopping mall environments," in *Proc. IEEE International Conference on Communications Workshops (ICCW)*, May 2016.

[22] D. Chizhik, J Ling, R. A. Valenzuela, "Self-alignment of interference arising from hallway guidance of diffuse fields", *IEEE Trans. on Wireless Communications*, v. 13, no. 7, pp.3853 - 3862, July, 2014.

[23] T. S. Rappaport, Y. Xing, G. R. MacCartney, A. F. Molisch, E. Mellios and J. Zhang, "Overview of millimeter wave communications for fifth-generation (5G) wireless networks-with a focus on propagation models," *IEEE Trans. on Antennas and Propagation*, v. 65, no. 12, pp. 6213-6230, Dec. 2017.

[24] "Propagation data and prediction methods for the planning of short-range outdoor radio communication systems and radio local area networks in the frequency range 300 MHz to 100 GHz," Recommendation ITU-R P.1441-9, Jun. 2017.

[25] A. Karttunen, A. F. Molisch, S. Hur, J. Park, C. J. Zhang ,"Spatially Consistent Street-by-Street Path Loss Model for 28-GHz Channels in Micro Cell Urban Environments", *IEEE Trans. on Wireless Comm.*, vol.16, no. 11, pp. 7538 – 7550, 2017

[26] Dai Lu and D. Rutledge, "Investigation of indoor radio channels from 2.4 GHz to 24 GHz," in proc. *IEEE Antennas and Propagation Society International Symposium*, 2003, pp. 134-137.

[27] T. S. Rappaport, G. R. MacCartney, S. Sun, H. Yan and S. Deng, "Small-Scale, Local Area, and Transitional Millimeter Wave Propagation for 5G Communications," *IEEE Transactions on Antennas and Propagation*, vol. 65, pp. 6474-6490, Dec. 2017.

[28] W. C. Jakes, Ed., Microwave Mobile Communications. New York, Wiley, 1974.

[29] A., Dvoretzky; Kiefer, J.; Wolfowitz, J. (1956). "Asymptotic minimax character of the sample distribution function and of the classical multinomial estimator". *Annals of Mathematical Statistics*. 27 (3): 642–669.
[30] W. B. Westphal and A. Sils, “Dielectric Constant and Loss Data”, Massachusetts Institute of Technology, Technical Report AFML-TR-72-39 9, Apr. 1972.
[31] D. C. Montgomery and G. C. Runger, Applied Statistics and Probability for Engineers, 3rd edition, Wiley, 2003.